# Stress peaks and rupture in star-polymer networks with bidisperse arm lengths

Yuichi Masubuchi*

*Department of Materials Physics, Nagoya University, Nagoya 464-8603, Japan*

* To whom correspondence should be addressed
mas@mp.pse.nagoya-u.ac.jp

## Abstract

The tensile response of phantom star-polymer networks with bidisperse arm lengths is examined by energy-minimization simulations. Short and long precursors form short–short, short–long, and long–long strands within one end-linked network. The contrast among these strands, composition, functionality, and reaction conversion are varied, and their effects on network rupture are examined. At a large length contrast, two stress peaks appear over part of the composition range. In such realizations, short–short strands carry most of the stress at the first peak, while short–long strands carry most of it after the intervening stress drop. Interesting findings are i) that the peak stretches, including those in double-peak cases, follow composition-dependent master curves when plotted against cycle-rank density, as in monodisperse networks, and ii) that a recently proposed mechanical model describes their cycle-rank dependence, with parameters that vary with composition and arm-length contrast. The parameters differ between the peak series, although their separate determination is limited in some conditions. In addition, the Mark-type improvement of the mechanical properties of the networks by adding a fraction of short strands is discussed. With the implemented FENE force law, the work up to the global stress peak is closer to the long-only value, but an improvement above that reference is not statistically resolved.

**Keywords**: gels; rubbers; fracture; rupture; coarse-grained simulations; molecular weight distributions

## Introduction

The molecular weight of network strands affects both deformation and rupture in gels and rubbers. For a network containing strands of a single molecular weight, the contour length provides a characteristic scale for extension [1–6]. However, rupture also depends on how the strands are connected and how deformation is distributed among them. When short and long strands coexist, changing their proportion changes not only the mean molecular weight but also the mechanical environment of each strand [7–16].

Networks with two strand-length populations have been investigated experimentally and numerically. Llorente et al.[11] prepared bimodal poly(dimethylsiloxane) networks by end-linking mixtures of long and very short chains. The short chains increased the ultimate strength without the usual decrease in extensibility, resulting in a larger work to rupture. Genesky et al.[16] subsequently combined experiments and simulations and associated the enhanced response with the limited extensibility of short chains rather than their aggregation. In contrast, Kondo et al.[14] found that the physical properties of the examined bimodal tetra-PEG gels above the precursor overlap concentration could be described using the average strand length. Thus, mixing two chain lengths does not always improve the mechanical properties.

Simulations provide detailed information on strand deformation[17] . Kamerlin and Elvingson[12] examined ideal bimodal networks and found increased extensibility and toughness relative to unimodal networks with the corresponding average chain length. A previous Brownian simulation study examined tetra-gel-type phantom networks made from stars containing both short and long arms[15]. Increasing the arm-length contrast increased double linkages and reduced the fraction of mechanically effective strands. However, fracture behavior did not correlate with the double-linkage fraction because scission occurred at single linkages. Arora[13] recently examined bimodal networks formed by connecting linear chains to multifunctional crosslinkers. Increasing the short-chain fraction reduced the ultimate stress and strain, although finite extensibility increased strain hardening at low and intermediate strains. Interactions omitted from a phantom model can also affect rupture. Furuya and Koga [18] simulated radical-polymerized networks and reported that entanglements among strands yield high toughness for some conditions. Hagita and Murashima[19] examined the combined effects of bond breakage and network heterogeneity on stress–strain hysteresis and scattering. These studies imply the need to distinguish strand-length effects from the effects of interactions, preparation, and scission rules.

Nevertheless, load transfer between components and the origin of network toughness in bimodal networks remain unclear. The present study examines the tensile response of bidisperse star-polymer networks. The rupture is analyzed for various prepolymer functionalities, conversions, compositions, and arm-length contrasts using a recently proposed mechanical model [20,21]. Stress contributions, strand scission, and periodic connectivity are also examined to identify the load-bearing strands. A comparison of Gaussian and finitely extensible springs tests whether short-precursor addition improves the tensile properties relative to a long-only network.

## Model and Simulations

The bead–spring model without excluded volume used previously was employed[22]. Each precursor star polymer consists of a central bead and $f$ arm chains with equal numbers of beads. Short and long precursors have $N_S = 2$ and $N_L = 8$, 16, or 32 beads per arm, respectively. Units of length, energy, and time are chosen as the average bond length $b$, thermal energy $k_BT$, and diffusion time of a single bead $\tau = \zeta b^2/k_BT$, where $\zeta$ is the friction constant. For the main series, each spring has the Gaussian potential $U_G(r) = 3r^2/2$ in simulation units. A limited comparison uses the Warner FENE potential $U_F(r) = -\frac{KR_0^2}{2}\ln\left(1 - \frac{r^2}{R_0^2}\right)$, with $K = 3$ and $R_0 = 2$. Thus, the two potentials have the same stiffness at small bond extension. The FENE spring coefficient is capped beyond $r = R_0\sqrt{0.9}$, but activation of this cap was not recorded. Two mutually reactive precursor species are used at equal stoichiometry, and end-linking is allowed only between different species. Consequently, primary loops formed by connecting two arms of the same precursor are excluded, whereas multiple connections between different precursors can occur.

The composition $\phi_L$ is the fraction of all precursor beads belonging to long-arm precursors, including the central beads. Since all beads have the same mass, this is a precursor weight fraction. If $n_S$ and $n_L$ are the total numbers of short and long precursors, $\phi_L = n_L(fN_L + 1)/\{n_S(fN_S + 1) + n_L(fN_L + 1)\}$. Integer precursor numbers give small deviations from the nominal compositions. This fraction is different from the precursor number fraction and from the number fractions of the resulting strand classes.

End-linking produces short-short, short-long, and long-long strands containing $2N_S + 1$, $N_S + N_L + 1$, and $2N_L + 1$ basic bonds, respectively. Strand classes are assigned from the initial network and are retained after scission. The Gaussian series contains $f = 3, 4, 6$, and 8 and reaction conversions $p = 0.60, 0.70, 0.80, 0.90$, and 0.95. The systems contain approximately 35,000 beads at a target bead number density of 8 in a cubic periodic box. Long-only networks are included for each $N_L$, and the short-only network is a common reference.

Precursor positions are chosen randomly in the periodic box, with initial bond lengths of unity and random bond directions. Equilibration is attained by Brownian dynamics, where the numerical integration is performed with a 2nd-order scheme[23] with a time step of 0.01, and the duration for equilibration is 1000.

End-linking is then enabled while Brownian dynamics continues. Reaction tests are performed at every $\tau$. An available A end can react with a B end within a distance of 0.5, with probability 0.1 per test. The reaction probabilities for A–A and B–B pairs are zero. No additional prohibition of multiple connections between different precursors is imposed.

Network configurations are saved when the reacted-end fraction first reaches each prescribed conversion. The values $p$ in this paper label these target conversions; the realized values can be slightly higher. Each saved network is minimized before extension. The same preparation protocol supplies the networks at different conversions, so these conversion points do not necessarily represent independently prepared ensembles.

The network is uniaxially extended under periodic boundary conditions and constant volume[24]. Each affine step increases the Hencky strain by $2 \times 10^{-4}$. At each fixed strain, the configuration is relaxed with FIRE 2.0 [25,26], using a maximum bead-force criterion of $10^{-4}$ and a limit of $10^5$ iterations. The integration uses unit mass and semi-implicit Euler steps. The initial FIRE step is selected automatically from the force, with a displacement scale of 0.05 and a stability cap; its maximum is ten times the initial step. The standard mixing parameters used here are $\alpha_0 = 0.1$, $N_{\mathrm{min}} = 5$, $f_{\mathrm{inc}} = 1.1$, $f_{\mathrm{dec}} = 0.5$, $f_{alpha} = 0.99$, and $N_{\mathrm{delay}} = 20$.

After energy minimization, the most extended eligible bond is cut if its squared length exceeds a critical value $b_c^2$. Bonds directly attached to branch beads are protected, and bonds in chains already flagged as broken are excluded from further cutting. The network is relaxed again, and cutting and relaxation are repeated until no eligible bond exceeds the threshold. Only then is the strain increased. The same deterministic cutting rule applies to Gaussian and FENE springs. The calculation stops when the elastic network loses percolation in the elongational direction. For both Gaussian and FENE springs, the cutting threshold is set to $b_c^2 = 1.5$.

The stress tensor is evaluated from the virial of all surviving basic bonds, including bonds attached to branch points. A strand-class contribution uses the same virial expression restricted to bonds assigned to that initial class. Surviving fragments of a cut strand are included in this sum, although their contributions to stress are negligible after energy minimization. The global stress-peak stretch $\lambda_b$ and stress $\sigma_b$ are obtained from each individual nominal-stress curve. Local peaks are detected without smoothing using a prominence threshold of 15% of the global maximum stress. For the peak-resolved Gaussian analysis, the first retained peak is denoted by $\lambda_{p1}$, and the highest subsequent retained peak by $\lambda_{p2}$. Thus, in a curve with more than two retained maxima, $\lambda_{p2}$ denotes the strongest later peak rather than necessarily the second local maximum. A single-peak curve contributes only to the first series. Peak-count sensitivity was examined at 10% and 20% prominence for the detailed composition comparison.

For Gaussian networks, at each set of $N_S$, $N_L$, $\phi_L$, $f$, and $p$, 16 different simulation runs are performed for most cases, whereas 32 realizations are employed for selected $N_L = 32$ conditions. The FENE comparison is restricted to $N_S = 2$, $N_L = 32$, $f = 4$, $p = 0.95$, and $0.70 \leq \phi_L \leq 1.00$ with 8 realizations for each condition. Gaussian and FENE runs with the same realization identifier at a given composition start from the same network, but different compositions are not paired.

## Results and Discussion

Figure 1(d) shows the nominal stress–stretch curves for the $(N_{\mathrm{S}}, N_{\mathrm{L}}) = (2,8)$ networks with $f = 4$, $p = 0.95$, and $\phi_{\mathrm{L}} = 0.45$. Here, $\phi_{\mathrm{L}}$ is the bead weight fraction of long-arm precursors, including their central beads. The curves have a single maximum, followed by loss of periodic connectivity in the stretching direction. Figure 1(e) shows that short–short and short–long strands carry most of the stress. In the illustrated realization, short–short strands carry approximately 55% of the stress at the maximum. Figure 1(f) shows that they also account for most of the early scission, although strands in all three classes have broken by the terminal state. Direct connectivity analysis finds that both the short–short subnetwork and the complete network span the stretching direction at the last saved preterminal state, but neither spans it at the terminal state. Both losses lie within this output interval; their relative timing is not resolved.

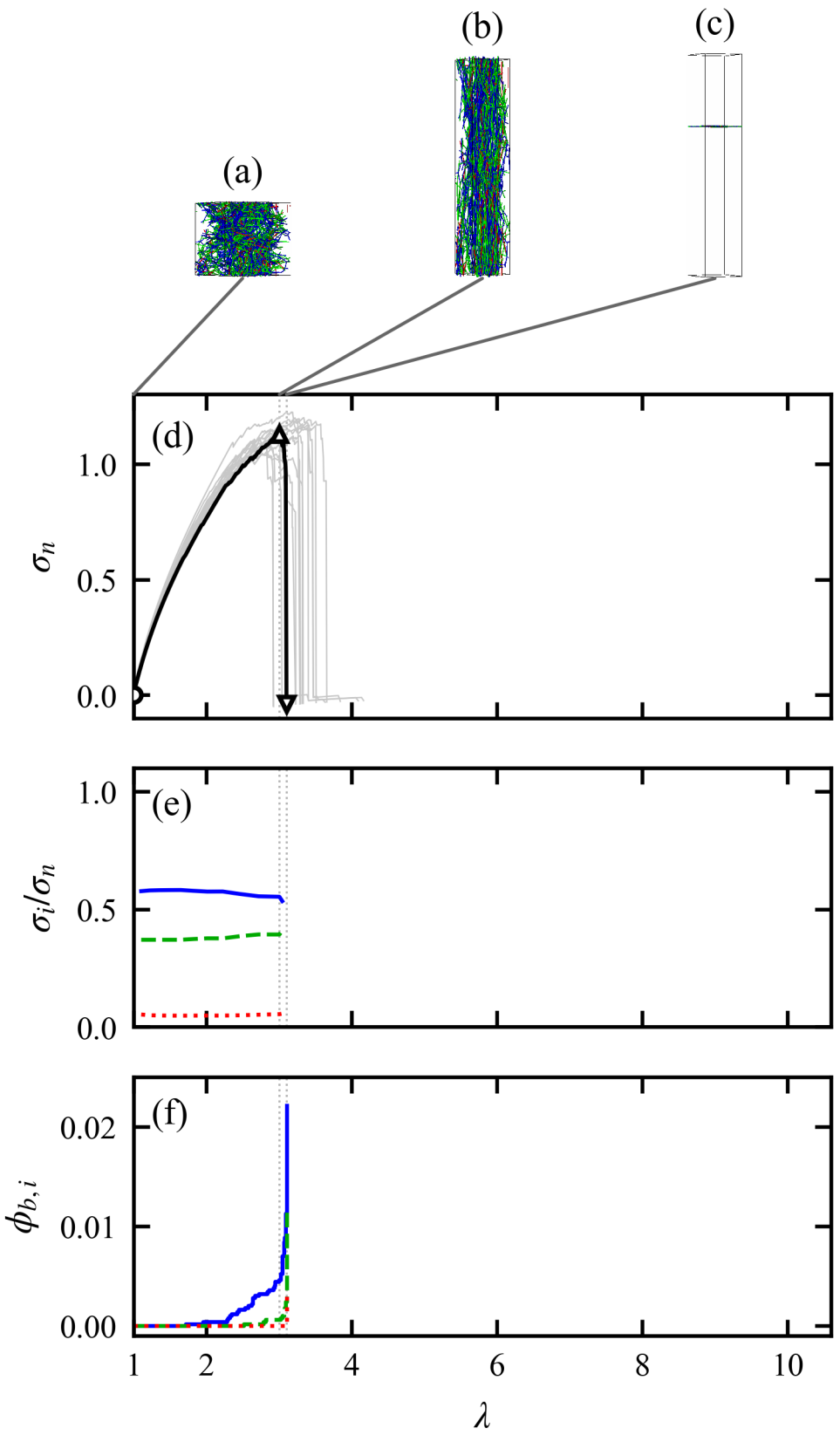

**Figure 1.** Rupture of $(N_{\mathrm{S}}, N_{\mathrm{L}}) = (2{,}8)$ networks with $f = 4$, $p = 0.95$, and $\phi_{\mathrm{L}} = 0.45$. Panels (a)–(c) show snapshots before stretching, at the stress maximum, and at the terminal state without spanning connectivity. Blue, green, and red indicate short–short, short–long, and long–long strands. Panel (d) shows nominal stress–stretch curves. Gray curves represent individual realizations; the bold black curve is the realization used for the snapshots and class-resolved analysis. Panels (e) and (f) show stress fractions and cut-strand counts divided by $N_{\mathrm{t}} {\sim} 5000$, the initial strand count. Solid blue, dashed green, and dotted red curves identify the three classes.

For the $(N_{\mathrm{S}}, N_{\mathrm{L}}) = (2{,}32)$ networks, two maxima occur over an intermediate range of composition. Figure 2(e) shows an example at $f = 4$, $p = 0.95$, and $\phi_{\mathrm{L}} = 0.45$. The stress of the illustrated realization reaches a first maximum at $\lambda \approx 2.2$, drops to a minimum at $\lambda \approx 2.3$, and then reaches a second maximum at $\lambda \approx 4.1$. In this realization, the second maximum is lower than the first. Here, $\lambda_{\mathrm{b}}$ is defined as the stretch at the global maximum of nominal stress. The first and second peak stretches are denoted by $\lambda_{\mathrm{p1}}$ and $\lambda_{\mathrm{p2}}$. These peak stretches are distinct from $\lambda_{\mathrm{perc}}$, the stretch at which periodic connectivity is lost in the stretching direction. The gray curves show the variation among realizations.

Figure 2(f) shows how the stress is shared among the strand classes. At the first peak, short–short strands carry 80% of the nominal stress. Their contribution drops to 1.2% at the subsequent minimum, while the short–long contribution rises to 97%. The blue strands contract between Figs. 2(b) and 2(c). Direct periodic-connectivity analysis confirms that the short–short subnetwork spans the stretching direction at the first peak but not at the minimum, while the complete network remains connected. At the minimum, ca. 2% of initial short–short strands have been cut. Unlike the single-peak realization in Fig. 1, the short–short subnetwork has therefore lost spanning connectivity while the complete network remains connected through the second peak. Short–long strands then carry most of the stress up to the second maximum; in this realization, their scission begins afterward. Thus, the second peak reflects loading and subsequent failure of the remaining network. The relative contribution of long–long strands becomes appreciable only later, when the total stress is small.

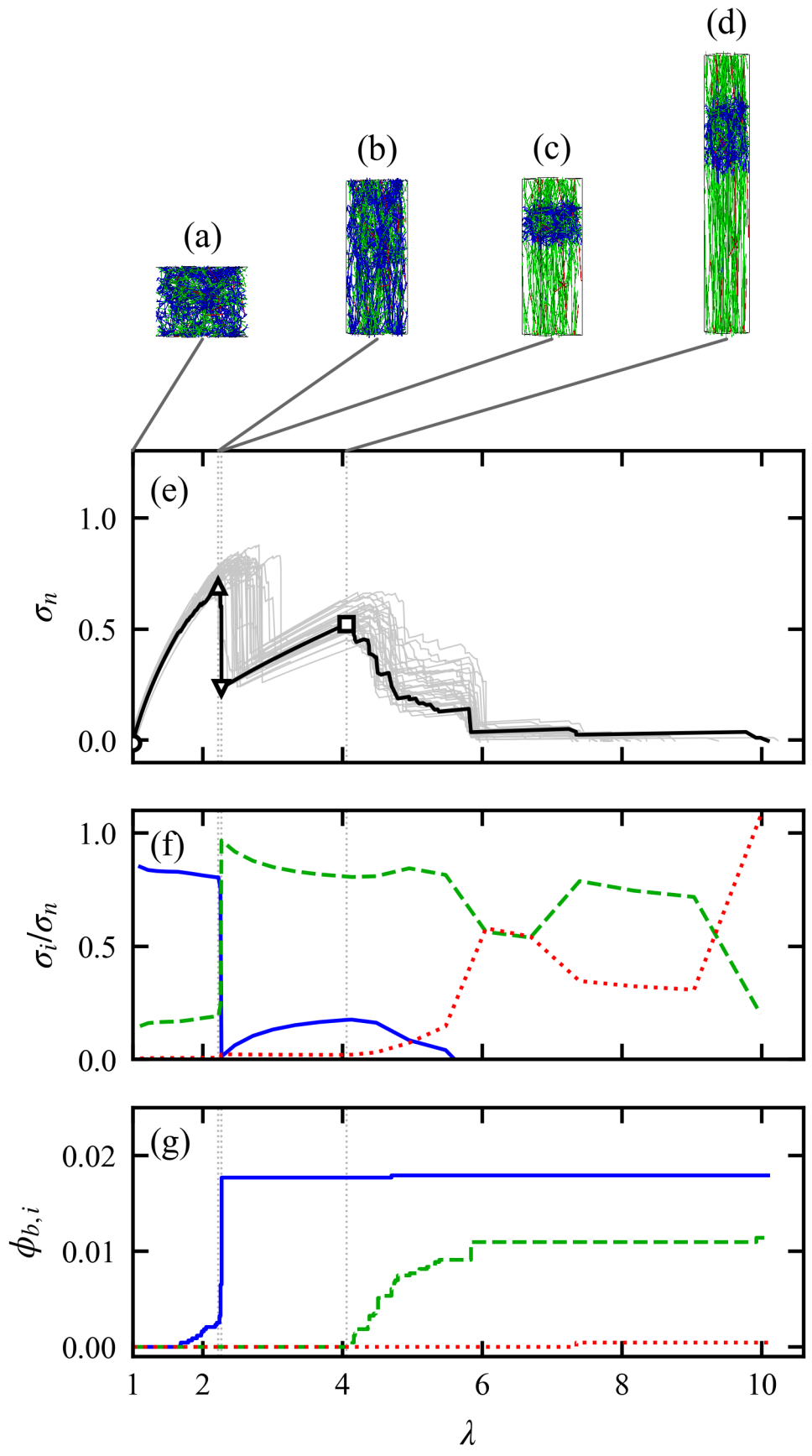

**Figure 2.** Rupture of $(N_S, N_L) = (2,32)$ networks with $f = 4$, $p = 0.95$, and $\phi_L = 0.45$. Panels (a)–(d) show snapshots before stretching, at the first stress maximum, at the following minimum, and at the second maximum. Strand colors are as in Fig. 1. Panel (e) shows nominal stress–stretch curves; the bold black realization supplies the snapshots and class-resolved analysis. Panels (f) and (g) show stress fractions and cut-strand counts divided by the initial total $N_t \sim 4300$. Solid blue, dashed green, and dotted red curves represent short–short, short–long, and long–long strands.

The second maximum in a periodic, strain-controlled calculation may not be reached in a macroscopic tensile test. The stress drop after the first maximum could promote strain localization or growth of a pre-existing crack, terminating the test before the remaining material reaches a second maximum. This possibility cannot be examined by the present homogeneous periodic-cell calculations. Previous Brownian simulations of bidisperse phantom networks[15] and the bimodal-network calculations of Arora[13] provide useful comparisons, but differences in relaxation and scission protocols must be distinguished from differences in strand-length distribution. Finite-rate relaxation or stochastic scission could alter the separation of the two loading stages; their effects cannot be inferred from the present quasistatic results alone.

Tauber et al. [27] provide a related example in double-network simulations, where stress redistribution changes the damage process relative to an affine prediction. The common feature is load transfer after early damage. Their separately prepared, interpenetrating networks, however, differ from the present strand classes, which share junctions within a single phantom network. In particular, a short–long strand joins short and long precursor arms and has no counterpart as an independent matrix network. The present composition therefore cannot be identified with the prestretch of a sacrificial network. The comparison concerns load redistribution, rather than equivalence of the two network architectures. Orr et al. [28] also related the enhanced toughness of double networks to spatially distributed rearrangements and stress redistribution. They prepared their networks sequentially by swelling a first network and polymerizing a second, whereas the present short and long strands belong to a single end-linked network.

The peak stretches are next examined using the mechanical model[20], which considers a bottleneck strand and the surrounding medium by elements connected in series. When the bottleneck reaches its critical stretch $\lambda_{\mathrm{bs}}$, the model relates the corresponding macroscopic stretch $\lambda_{\mathrm{b}}$ to the cycle-rank density as follows.

$$\lambda_b - 1 = (\lambda_{bs} - 1)\left[1 + \frac{\nu_h}{1 + \alpha\,\xi}\right] \qquad (1)$$

Here, $\nu_{\mathrm{h}}$ represents the number of medium units in series with the bottleneck, and $\alpha$ describes redundancy. Following the previous study[20], $\alpha$ is fixed at 2 because $\alpha$ and $\nu_{\mathrm{h}}$ are difficult to determine independently.

The graph-based cycle-rank density[29–32] is defined by Eq. (2).

$$\xi = \frac{1}{n_n}(n_s - n_n) \qquad (2)$$

Here, $n_s$ and $n_n$ count strands and their incident nodes in the reduced backbone graph extracted from each initial, uncut network. No additional winding selection or degree-two contraction is applied when these counts are evaluated. The resulting $\xi$ approaches the Miller–Macosko value[33,34] at high conversion, but differences remain at lower conversion. Fits use condition means for $\xi \geq 0.1$, weighted by the inverse squared standard error of the mean peak stretch, with a minimum standard error of 0.01. At least eight realizations per condition and four fitted conditions are required. The condition-mean $\xi$ is treated as fixed; its variation within each condition and covariance between conversions sharing a preparation trajectory are not propagated into the parameter errors.

Figure 3(a) shows $\lambda_{\mathrm{b}}$ against $\xi$ for the $(N_{\mathrm{S}}, N_{\mathrm{L}}) = (2,8)$ networks. The long and short references (shown in ochre and gray) draw master curves, as reported for a variety of systems composed of monodisperse strand lengths[6,35–40]. Notably, the data lie on master curves regardless of mixing. Further, the curve for each composition is approximately described by Eq. (1) with composition-dependent parameters. Deviations remain toward lower connectivity. Figures 3(b) and 3(c) show an overall increase in $\lambda_{\mathrm{bs}}$ and decrease in $\nu_{\mathrm{h}}$ as $\phi_{\mathrm{L}}$ increases, with smaller non-monotonic variations. This trend connects the short-only and long-only limits and is consistent with a gradual change in the strands carrying load.

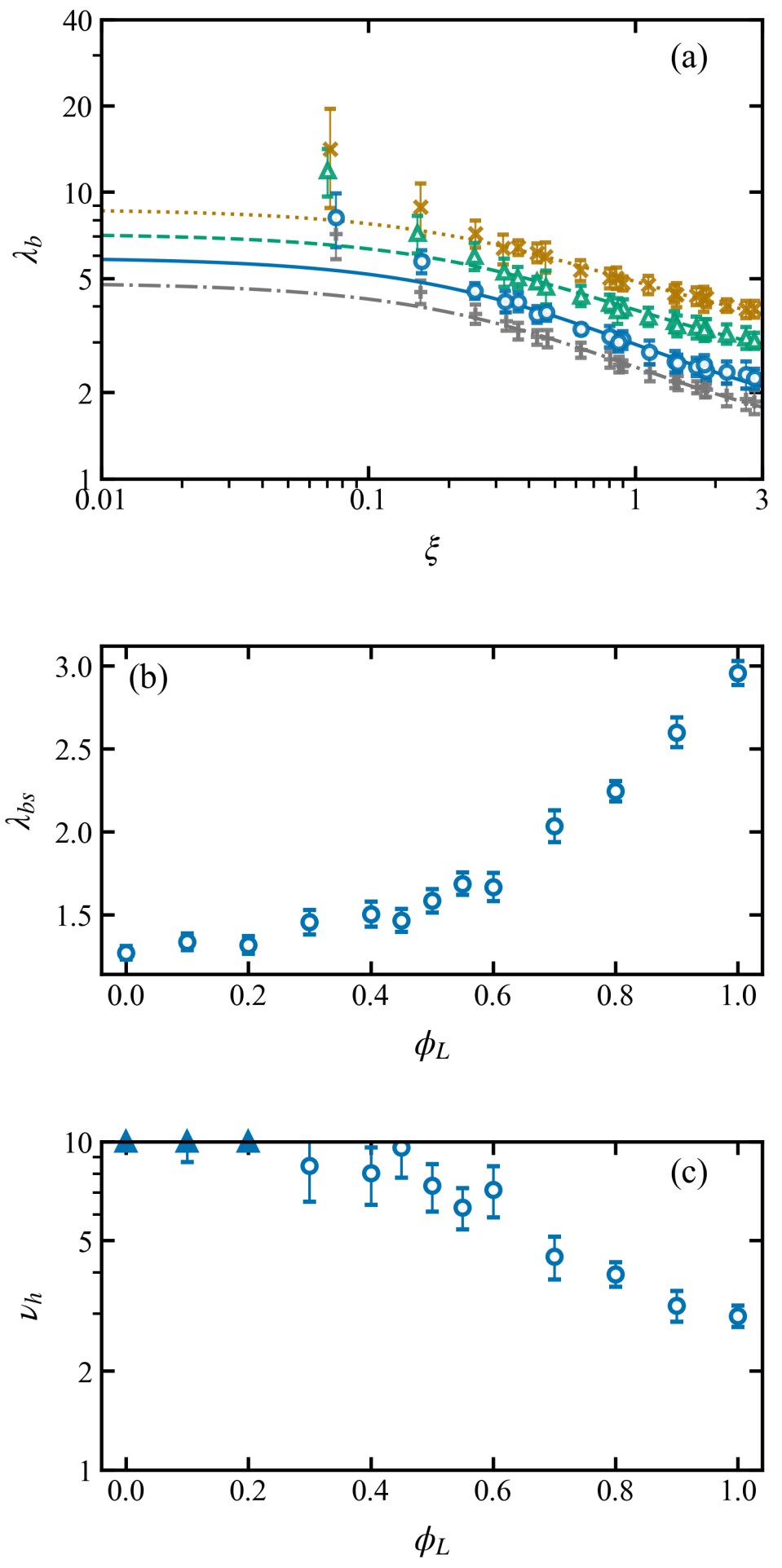

**Figure 3.** Mechanical-model analysis for $(N_S, N_L) = (2,8)$. (a) Condition-mean $\lambda_b$ versus snapshot-derived $\xi$ for short-only networks (gray crosses, dash-dotted line), long-only networks (ochre crosses, dotted line), $\phi_L = 0.45$ (blue circles, solid line), and $\phi_L = 0.80$ (green triangles, dashed line). Conditions are $f = 3,4,6,8$ and $p = 0.60,0.70,0.80,0.90,0.95$; error bars show realization standard deviations. The $\xi$ and stretch means use the same realizations. Curves are fitted for $\xi \geq 0.1$ and extended down to 0.01. (b) Fitted $\lambda_{bs}$ and (c) $\nu_h$ versus long-precursor weight fraction. Parameter bars show local standard errors, with covariance scaled by the reduced chi-squared; covariance between conversions is not included. Upward triangles at the top edge mark estimates of 13.2, 10.7, and 12.6 at $\phi_L = 0$, 0.10, and 0.20; some error bars also extend above the frame. The global stress maximum defines $\lambda_b$, including realizations with secondary maxima.

Figure 4 shows the corresponding results for $(N_{\mathrm{S}}, N_{\mathrm{L}}) = (2,16)$. The global-peak stretches are again approximately described by Eq. (1) over the fitted $\xi$ range. The composition dependence differs from that in Fig. 3: $\lambda_{\mathrm{bs}}$ decreases slightly and $\nu_{\mathrm{h}}$ increases up to approximately $\phi_{\mathrm{L}} = 0.4$, followed by an increase in $\lambda_{\mathrm{bs}}$ and a decrease in $\nu_{\mathrm{h}}$. These changes suggest a crossover in the effective load-bearing response. As for Fig. 3, the definition of $\lambda_{\mathrm{b}}$ remains the global maximum, including realizations with secondary maxima.

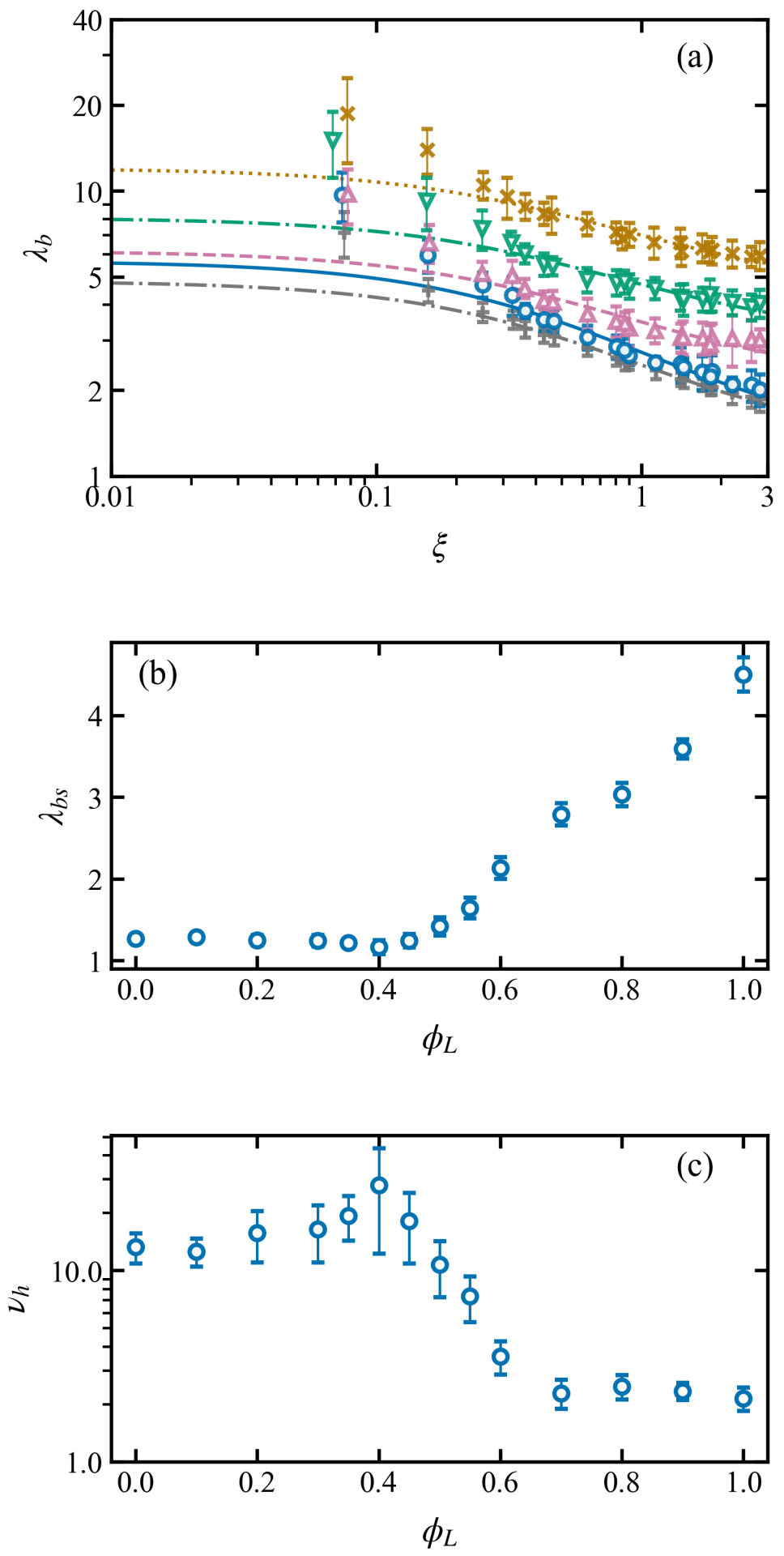

**Figure 4.** Mechanical-model analysis for $(N_S, N_L) = (2{,}16)$. (a) Condition-mean $\lambda_b$ versus snapshot-derived $\xi$. The short-only and long-only references are as in Fig. 3; blue circles and a solid line show $\phi_L = 0.45$, purple upward triangles and a dashed line show 0.60, and green downward triangles and a dash-dotted line show 0.80. Conditions, fitting range, and realization-standard-deviation bars are as in Fig. 3. (b) Fitted $\lambda_{bs}$ and (c) $\nu_h$ versus long-precursor weight fraction. Parameter bars and averaging are as in Fig. 3. The global stress maximum defines $\lambda_b$ throughout.

For $(N_S, N_L) = (2{,}32)$, Eq. (1) is applied separately to $\lambda_{p1}$ and $\lambda_{p2}$. Figure 5(a) shows the two peak series at $\phi_L = 0.45$ together with the short-only and long-only controls. The additional compositions $\phi_L = 0.10$ and 0.90 show the change from a short-dominated to a long-dominated first-peak response. Single-peak realizations contribute to the first-peak series, whereas the second-peak series contains only realizations with a detected second maximum. The first-peak stretches are close to the short-only response over part of the $\xi$ range, while the second-peak stretches lie below the long-only response. Figures 5(b) and 5(c) show the independently fitted parameters. At $\phi_L = 0.55$ and 0.60, the first-series fits reach the $\lambda_{bs} = 1$ boundary, and no finite $\nu_h$ can be determined. The finite fitted quantity is the product $(\lambda_{bs} - 1)\nu_h$ in this limiting form; it should not be interpreted as a directly measured number of medium strands. For the second series, $\lambda_{bs}$ increases overall from about 2.7 to 3.4 between $\phi_L = 0.20$ and 0.60, whereas $\nu_h$ varies non-monotonically between about 1.6 and 2.2. The two parameters are strongly anticorrelated. Their fitted values summarize the peak response but cannot be assigned independently to a number or stress weight of load paths.

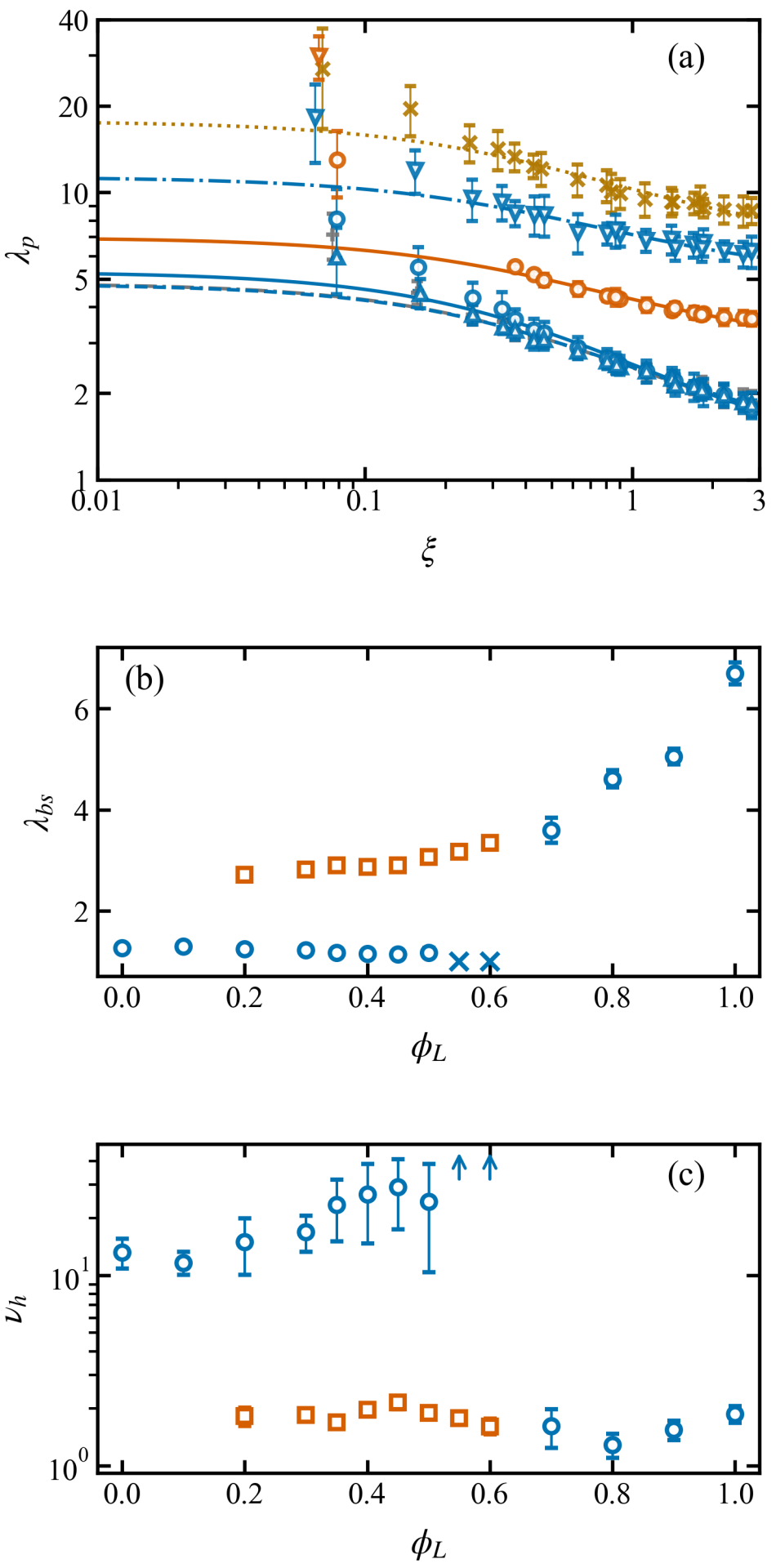

**Figure 5.** Mechanical-model analysis for $(N_S, N_L) = (2{,}32)$. (a) Condition-mean peak stretches versus snapshot-derived $\xi$. Gray and ochre crosses show short-only and long-only controls; blue and orange identify first and second peaks. Circles and solid curves show $\phi_L = 0.45$, upward triangles and a dashed curve show 0.10, and downward triangles and a dash-dotted curve show 0.90. No second-peak condition at 0.10 meets the minimum realization count. At 0.90, one second-peak condition is shown below $\xi = 0.1$, but no second-series fit is made. Conditions, fitting range, and realization-standard-deviation bars are as in Fig. 3. (b) Fitted $\lambda_{bs}$ and (c) $\nu_h$, with parameter bars and averaging as in Fig. 3. Blue circles and orange squares identify the first and second series. Single-peak realizations enter the first series; only detected second peaks enter the second. The $\nu_h$ axis is logarithmic. Crosses at $\lambda_{bs} = 1$ and upward arrows in (c) mark limiting first-series fits at $\phi_L = 0.55$ and 0.60, where no finite $\nu_h$ is determined. The arrows are not numerical lower bounds.

The results in Figs. 1–5 distinguish a gradual change in the global response at smaller chain-length contrast from a separation of loading stages at larger contrast. In the two-peak example, the short–short subnetwork stops spanning the stretching direction between the first peak and the following minimum, while the complete network remains connected. The load is then carried mainly by short–long strands until a later stretch. The mechanical-model parameters nicely capture the changes in peak stretch with connectivity and composition.

The effect of adding short strands is finally examined in terms of the ultimate properties relative to a long-only network. For instance, Llorente et al.[11] reported increased strength without the usual loss of extensibility in bimodal PDMS networks, giving a larger work to rupture. Figure 6 compares the global-peak stretch $\lambda_\mathrm{b}$, maximum nominal stress $\sigma_\mathrm{b}$, and work $W_\mathrm{b}$ for the present $(N_\mathrm{S}, N_\mathrm{L}) = (2{,}32)$ networks over $0.70 \le \phi_\mathrm{L} \le 1.00$. Each quantity is normalized by the mean for the corresponding long-only spring model. The peak-limited work is defined by Eq. (3).

$$W_b = \int_1^{\lambda_b} \sigma_n(\lambda)\, \mathrm{d}\lambda \qquad (3)$$

Adding short strands increases the maximum stress but reduces the global-peak stretch in both spring models. For Gaussian springs, the peak-limited work remains below the long-only value over the compositions in Fig. 6. The FENE calculation gives larger work and normalized values closer to unity. For example, at $\phi_\mathrm{L} = 0.80$, the average FENE work ratio is 0.98, although some realizations perform better.

Nevertheless, the present data do not establish an improvement over the FENE long-only control, and the mechanism remains open. The comparison with experiments requires attention to the integration limit: $W_\mathrm{b}$ ends at the global stress maximum, whereas experimental work to rupture can include a later part of the tensile curve. Work integrated to loss of periodic connectivity is a separate observable, but its mean also remains below the long-only value for these FENE mixtures. The classical PDMS networks had a larger chain-length contrast than the networks studied here. For Llorente et al.[11], the molecular weights of the long and short chains were 18500 and either 660 or 220 g mol$^{-1}$, corresponding to ratios of approximately 28 and 84. The largest precursor-arm ratio examined here is 16, and the ratio of the longest to shortest actual strand lengths is 65/5=13. A larger contrast could change the separation of the loading stages and the effect of short-chain addition.

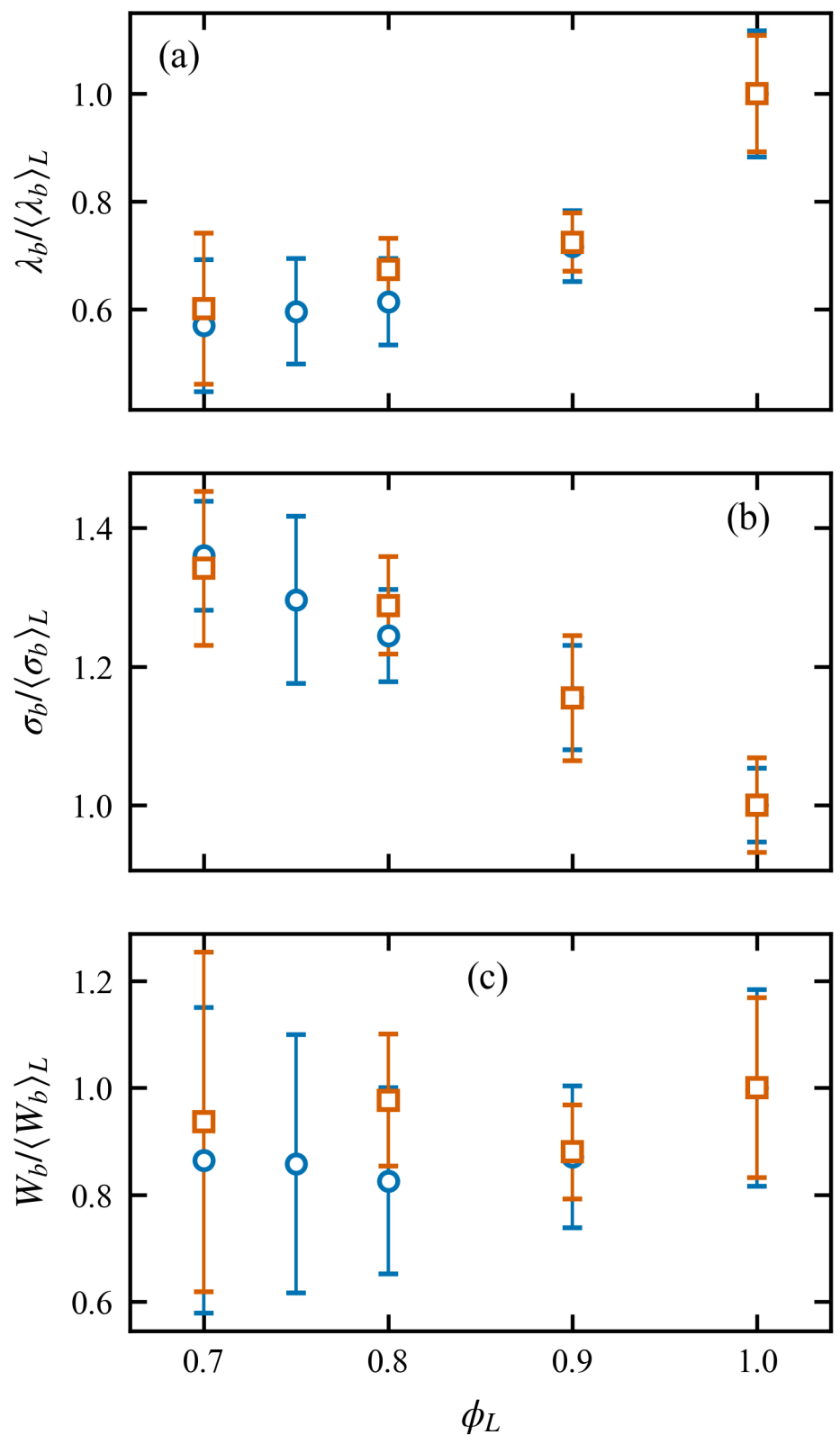

**Figure 6.** Long-only-normalized (a) global-peak stretch $\lambda_\mathrm{b}$, (b) maximum nominal stress $\sigma_\mathrm{b}$, and (c) peak-limited work $W_\mathrm{b}$ for $(N_\mathrm{S}, N_\mathrm{L}) = (2{,}32)$, $f = 4$, and $p = 0.95$ over $0.70 \le \phi_\mathrm{L} \le 1.00$. Blue circles show Gaussian springs (16 realizations per composition), and orange squares show FENE springs (eight realizations). Symbols show means; bars show realization standard deviations divided by the mean of the corresponding spring model at $\phi_\mathrm{L} = 1$. Denominator uncertainty is not propagated into these bars. No FENE point is available at $\phi_\mathrm{L} = 0.75$. Work is integrated from $\lambda = 1$ to the global stress maximum, not to loss of connectivity.

## Conclusions

Phantom-chain simulations were performed for networks formed by end-linking short-arm and long-arm star precursors. The stress response changes with both composition and arm-length contrast. For the large contrast $N_S = 2$ and $N_L = 32$, two prominent stress peaks occur over part of the composition range. Strand-resolved analysis shows that early damage to short-short strands can be followed by transfer of the main stress contribution to short-long strands. In the representative two-peak realization, short-short connectivity is lost while the whole network still spans the stretching direction.

The peak stretches, including those for both single- and double-peak cases, can be organized by cycle rank, with composition-dependent master curves similar to those reported for monodisperse precursor networks. These curves were analyzed with the recently reported mechanical model. The fitted $\lambda_{bs}$ and $\nu_h$ depend on composition and length contrast, and the two peak series require independent parameter estimates. These parameters effectively describe the peak positions, although their covariance limits separate structural interpretation in poorly determined cases.

Concerning the experimentally established improvement of the mechanical properties by the inclusion of short strands, the Gaussian networks exhibit a rise in peak stress but a drop in peak stretch, and the work up to the global stress peak decreases relative to the long-only reference. Using the implemented FENE force law in place of Gaussian springs increases the peak-limited work relative to the Gaussian result and brings it closer to the reference. Nevertheless, a statistically supported increase is not established. Larger segment length contrasts and interactions absent from the phantom model remain relevant when comparing with the reinforcement reported for bimodal PDMS networks.

## Acknowledgements

This study was partly supported by the Eno Science Foundation and JSPS KAKENHI (26H02291).